\documentstyle[12pt,epsf]{article}

\newcommand{\be}{\begin{equation}}
\newcommand{\ee}{\end{equation}}
\newcommand{\bea}{\begin{eqnarray}}
\newcommand{\eea}{\end{eqnarray}}
\newcommand{\ba}{\begin{array}}
\newcommand{\ea}{\end{array}}
\newcommand{\bean}{\begin{eqnarray*}}
\newcommand{\eean}{\end{eqnarray*}}
\newcommand{\ct}{\cite}
\newcommand{\eq}[1]{(\ref{#1})}
\newcommand{\n}{\nonumber}
\newcommand{\vs}{\vspace}
\def\l{\label}
\def\a{\alpha}

\def\d{\delta}

\def\pa{\partial}

\def\ra{\rightarrow}

\def\ca{{\cal A}}

\def\ch{{\cal H}}

\def\4r{{\bf R}^4}
\makeatletter
\begin{document}

\topmargin 0pt

\oddsidemargin -3.5mm

\headheight 0pt

\topskip 0mm \addtolength{\baselineskip}{0.20\baselineskip}

\renewcommand{\thefootnote}{\fnsymbol{footnote}}

\begin{flushright}
SOGANG-HEP 269/00 \\
{\tt hep-th/0003093}
\end{flushright}
\vs{5mm}
\begin{center}
{\large \bf Comments on Instantons on Noncommutative $\4r$}\\
\vs{5mm} Keun-Young Kim, Bum-Hoon Lee and Hyun Seok Yang
\footnote{Present address: Department of Physics, National Taiwan University,
Taipei 106, Taiwan, R.O.C.}\\
\vs{5mm}
{\it Department of Physics, Sogang University, Seoul 121-742, Korea}\\
\vs{5mm}
{\it E-mail:} {\tt kykim@physics4.sogang.ac.kr,
bhl@ccs.sogang.ac.kr,
hsyang@phys.ntu.edu.tw}\\
\vs{5cm}

{\bf ABSTRACT}
\end{center}
We study $U(1)$ and $U(2)$ instanton solutions on noncommutative
$\bf{R}^4$ based on the noncommutative version of ADHM equation
proposed by Nekrasov and Schwarz. It is shown that the
anti-self-dual gauge fields on self-dual noncommutative $\4r$
correctly give integer instanton numbers for all cases we
consider. We also show that the completeness relation in the ADHM
construction is generally satisfied even for noncommutative
spaces.

\vs{1.5cm}
\begin{flushleft}
\today \\
\end{flushleft}

\newpage

\renewcommand{\thefootnote}{\arabic{footnote}}
\setcounter{footnote}{0}

\section{Introduction}

Recently it has been known \ct{CDS,DH,SW} that quantum field
theory on noncommutative space can arise naturally as a decoupled
limit of open string dynamics on D-branes in the background of
Neveu-Schwarz two-form field $B_{NS}$. In particular, it was shown
in \ct{CDS,DH} that noncommutative geometry can be successfully
applied to the compactification of M(atrix) theory \ct{BFSS,IKKT}
in a certain background and the low energy effective theory for
D-branes in the $B_{NS}$ field background, which are specifically
described by a gauge theory on noncommutative space \ct{SW}.

In their paper \ct{NS}, Nekrasov and Schwarz showed that
instanton solutions in noncommutative Yang-Mills theory can be
obtained by Atiyah-Drinfeld-Hitchin-Manin (ADHM) equation
\ct{ADHM} defined on noncommutative $\4r$ which is equivalent to
adding a Fayet-Iliopoulos term to the usual ADHM equation. The
remarkable fact is that the deformation of the ADHM equation has
an effect removing the singularity of the instanton moduli space
\ct{NS,MB,HN,BN,KF,Ho}.

In this report, we study $U(1)$ and $U(2)$ instanton solutions on
noncommutative $\bf{R}^4$ based on the noncommutative version of
ADHM equation proposed by Nekrasov and Schwarz. Here it is shown
that the anti-self-dual gauge fields on self-dual noncommutative
$\4r$ correctly give integer instanton numbers for all cases we
consider. After this paper, in \ct{corr}, a method to obtain
(anti-)self-dual solutions with integer instanton number was
proposed using 't Hooft ansatz although the resulting field
strength is not Hermitian.

The paper is organized as follows. In next section we review the
ADHM construction of noncommutative instantons on self-dual
noncommutative $\4r$. In section 3 we explicitly calculate the
anti-self-dual field strengths for simple cases, $k=1$ and $k=2$
for $U(1)$ and $k=1$ for $U(2)$. We show that these solutions
correctly give integer instanton numbers. In section 4 we discuss
the completeness relation as an ADHM condition. It is also shown
that the completeness relation in the ADHM construction is
generally satisfied even for noncommutative spaces. In section 5
we discuss the results obtained and address some issues.

Recently, in their paper \ct{CKT}, Chu, et al. pointed out that
the topological charge of noncommutative instantons by ADHM
construction we considered in this paper is correctly an integer.
According to their pointing out, we have redone the numerical
calculation on the instanton number using Maple. Now we have
obtained integer instanton numbers for all solutions previously
found in this paper. It turned out that the previous puzzling
result had been caused by an error of our numerical calculation.

\section{Noncommutative Version of ADHM Equation}

Noncommutative $\4r$ is described by algebra generated by $x^\mu$
obeying the commutation relation: \be \l{crR4} [x^\mu,
x^\nu]=i\theta^{\mu\nu}, \ee where $\theta^{\mu\nu}$ is a
non-degenerate matrix of real and constant numbers. Since we are
interested in noncommutative instanton backgrounds and the
instanton moduli space only depends on the self-dual part
$\theta^+ = 1/2(\theta + * \theta)$ \ct{SW,NS}, we restrict
ourselves to the case where $\theta^{\mu\nu}$ is self-dual and set
$$\theta^{12}=\theta^{34}=\frac{\zeta}{4}.$$
Then the algebra in \eq{crR4}, denoted as $\ca_{\zeta}$, depends
only on one parameter $\zeta$ where we choose $\zeta > 0$. We
consider an algebra ${\cal A}_{\zeta}$ consisted of smooth
operators ${\cal O}$. The commutation relations (\ref{crR4}) have
an automorphism of the form $ x^{\mu} \mapsto x^{\mu} + c^{\mu}$,
where $c^{\mu}$ is a commuting real number, and we denote the Lie
algebra of this group by ${\bf \underline{g}}$. Since the
derivative operator $\pa_{\mu}$ can be understood as the action of
${\bf \underline{g}}$ on ${\cal A}_{\zeta}$ by translation, the
generators of ${\bf \underline{g}}$ can be defined by unitary
operators $U_c=e^{c^\mu \pa_\mu}$ where
$\pa_\mu=-iB_{\mu\nu}x^\nu$ with $B_{\mu\nu}$, an inverse matrix
of $\theta^{\mu\nu}$. Then the derivative for a smooth operator
${\cal O}$ is defined by $\partial_\mu {\cal O} \equiv
[\partial_\mu,{\cal O} ]$. One can check that $\pa_\mu$ satisfies
following commutation relations
\[[\partial_\mu,x^\nu]=\d_\mu^\nu, \quad
[\partial_\mu,\partial_\nu]=iB_{\mu\nu}.\]
However the action of two derivatives on any operator ${\cal O}$
commutes:
$$\pa_\mu\pa_\nu {\cal O}- \pa_\nu\pa_\mu {\cal O}=
[[\pa_\mu,\pa_\nu], {\cal O}]=0.$$

Introduce the generators of noncommutative ${\bf C}^2 \approx
{\bf R}^4$ by \be z_1 = x^2 + i x^1, \quad z_2 = x^4 + i x^3. \ee
Their non-vanishing commutation relations reduce to \be
\label{noncom}
 [\bar{z}_1, z_1]
=[\bar{z}_2, z_2] =\frac{\zeta}{2}. \ee The commuatation algebra
is that of the annihilation and creation operators for a simple
harmonic oscillator (SHO) and so one may use the SHO Hilbert
space as a representation of $\ca_{\zeta}$ as adopted in
\ct{NS,KF}. Therefore let's start with the algebra End ${\cal H}$
with finite norm of operators acting on the Fock space ${\cal H} =
\sum_{(n_1,n_2) \in {\bf Z}_{\geq 0}^2 } {\bf C} | n_1 , n_2>$,
where $\bar{z}, z$ are represented as annihilation and creation
operators: \bea \l{sho} &&\sqrt{\frac{2}{\zeta} } \bar{z}_1 |
n_1, n_2> = \sqrt{n_1} | n_1 - 1, n_2>, \quad
\sqrt{\frac{2}{\zeta} } z_1 | n_1, n_2>
= \sqrt{n_1+1 } | n_1 +1, n_2>, \n\\
&&\sqrt{\frac{2}{\zeta} } \bar{z}_2 | n_1, n_2>
= \sqrt{n_2} | n_1, n_2-1>, \quad
\sqrt{\frac{2}{\zeta} } z_2 | n_1, n_2>
= \sqrt{n_2+1} | n_1, n_2+1>.
\eea

ADHM construction describes a way for finding anti-self-dual
configurations of the gauge field in terms of some quadratic
matrix equations on $\4r$ \cite{ADHM}. Recently, N. Nekrasov and
A. Schwarz made the ADHM construction to be applied to the
noncommutative ${\bf R}^4$ \cite{NS}. In order to describe $k$
instantons with gauge group
$U(N)$, one starts with the following data:\\
1. A pair of complex hermitian vector spaces $V={\bf C}^k,\;W={\bf C}^N$.\\
2. The operators $B_1, B_2 \in Hom(V,V),\;I \in Hom(W,V)$ and
$J \in Hom(V,W)$ satisfying the equations
\bea
\l{ADHM1}
&&\mu_r = [B_1 , B_1 ^\dagger ] +[B_2 , B_2 ^\dagger ] + II^\dagger -
J^\dagger J=\zeta,\\
\l{ADHM2}
&&\mu_c = [B_1 , B_2 ] + IJ=0.
\eea
3. Define a Dirac operator $D^\dagger : V \oplus V \oplus W
\rightarrow V \oplus V $ by
\begin{equation}
D^\dagger = \pmatrix { \tau_z \cr \sigma^\dagger_z }
\end{equation}
where
\begin{equation}
\tau_z = \pmatrix { B_2 - z_2 & B_1 - z_1 & I },
\quad \sigma_z = \pmatrix {-B_1 + z_1 \cr B_2 - z_2 \cr J }.
\end{equation}
Then the ADHM equations \eq{ADHM1} and \eq{ADHM2} are equivalent
to the set of equations \be \l{ADHM3}
\tau_z\tau^\dagger_z=\sigma^\dagger_z\sigma_z, \quad
\tau_z\sigma_z=0. \ee

According to the ADHM construction, we can get the gauge field
(instanton solution) by the formula
\begin{equation}
\l{gfA}
A_\mu = \psi^\dagger \partial_\mu \psi,
\end{equation}
where $\psi : W \rightarrow V \oplus V \oplus W $ is $N$
zero-modes of $D^\dagger$, i.e.,
\be \label{Firstadhm} D^\dagger
   \psi = 0. \ee
For given ADHM data and the zero mode condition \eq{Firstadhm},
the following completeness relation has to be satisfied to
construct an (anti-)self-dual field strength from the gauge field
\eq{gfA}
\begin{equation}\label{CR}
  D{1\over D^\dagger D}D^\dagger + \psi \psi^\dagger = 1.
\end{equation}
We will show in section 4 that this relation is always satisfied
even for noncommutative spaces.

Note that the vector bundle over the noncommutative space
$\ca_\zeta$ is a finitely generated projective
module.\footnote{The module ${\cal E}$ is projective if there
exists another module ${\cal F}$ such that the direct sum ${\cal
E} \oplus {\cal F}$ is free, i.e., ${\cal E} \oplus {\cal F}
\cong \ca_\zeta^N$ as right $\ca_\zeta$-module.} It was pointed
out in \ct{KF,Ho} that for a noncommutative space there can be a
projective module associated with a projection with non-constant
rank (so the corresponding bundle has non-constant dimension) and
the instanton module is the kind that is related to a projection
operator in $\ca_\zeta$ given by \be \l{p} p=\psi^\dagger \psi.
\ee Just as in the ordinary case, the anti-self-dual field
strength $F_A$ can be calculated by the following formula
\begin{eqnarray}
F_A&=&\psi^\dagger\left(d\tau_z^\dagger{1\over\triangle_z}d\tau_z
+d\sigma_z{1\over\triangle_z}d\sigma_z^\dagger\right)\psi
\nonumber \\
\label{Fformular}
&=& \psi^\dagger \pmatrix{dz_1{1\over\triangle_z}d\bar{z}_1
+ d\bar{z}_2{1\over\triangle_z}dz_2 &
  d\bar{z}_2{1\over\triangle_z}dz_1
- dz_1{1\over\triangle_z}d\bar{z}_2 &  0  \cr
  d\bar{z}_1{1\over\triangle_z}dz_2
- dz_2{1\over\triangle_z}d\bar{z}_1 &
  d\bar{z}_1{1\over\triangle_z}dz_1
+ dz_2{1\over\triangle_z}d\bar{z}_2 &  0  \cr
                           0 & 0 & 0 }    \psi,
\end{eqnarray}
where $\triangle_z=\tau_z\tau^\dagger_z=\sigma^\dagger_z\sigma_z$
has no zero-modes so it is invertible \ct{NS,KF}. Note that the
completeness relation \eq{CR} is used to derive the above field
strength $F_A$ which is anti-self-dual.

\section{Explicit Calculation of Instanton Charge}

In this section, we will perform an explicit calculation on the
instanton charge from the solutions obtained by the ADHM
construction in the previous section. First, we briefly do it for
single $U(1)$ instanton solution obtained by Nekrasov and Schwarz
\ct{NS} for the purpose of illuminating our calculational method.
And then we will do the same calculation for two $U(1)$
instantons and single $U(2)$ instanton solutions.

It is natural to require that the integration on a quantum, i.e.
noncommutative, space, which is the trace over its Hilbert space
or more precisely Dixmir trace, has to respect a symmetry $x \ra
x'=x'(x)$, that is, $$ \l{Tr} {\rm Tr}_{\cal H}\Bigl({\cal
O}(x')\Bigr)= {\rm Tr}_{\cal H}\Bigl({\cal O}(x)\Bigr). $$
Intuitively, the quantized $\4r$ in the basis \eq{sho} becomes
two dimensional integer lattice $\{(n_1, n_2) \in {\bf Z}_{\geq
0}^2\}$. Thus one can naturally think that the integration on
classical $\4r$ should be replaced by the sum over the lattice:
\be \l{trace} \int d^4x {\cal O}(x) \ra {\rm Tr}_\ch {\cal
O}(x)\equiv \Bigl(\frac{\zeta\pi}{2}\Bigr)^2 \sum_{(n_1,
n_2)}\langle n_1,n_2|{\cal O}(x)|n_1,n_2\rangle. \ee This
definition used in \ct{NS} to calculate the instanton number of
$U(1)$ solution indeed respects the translational symmetry as
well as the rotational symmetry since the automorphism ${\bf
\underline{g}}$ of $\4r$ such as \eq{Tr} can be generated by the
unitary transformation acting on $\ch$. If the system preserves
rotational symmetry, for example, single instanton solution, we
then expect for the case that the sum with respect to $(n_1,
n_2)$ can be reduced to that with respect to $N=n_1+n_2$
corresponding to a radial variable
$r=\sqrt{x_1^2+x_2^2+x_3^2+x_4^2}$.

\subsection{Single $U(1)$ Instanton}

In the ordinary case any regular $U(1)$ instanton solution can not
exist. However, in the noncommutative case, there are nontrivial
$U(1)$ instantons for every $k$ \ct{NS,KF,BN}. Suppose $(B_1,
B_2, I)$ is a solution to the equations \eq{ADHM1} and \eq{ADHM2}
where one can show $J=0$ for $U(1)$ \ct{HN}. If we write the
element of $V \oplus V \oplus W$ as $\psi = \psi_1 \oplus \psi_2
\oplus \xi$, Eq.\eq{Firstadhm} is then
\begin{equation}
\l{Dpsi}
D^\dagger\psi = \pmatrix { B_2 - z_2 & B_1 - z_1 & I \cr
              -B_1 ^\dagger + \bar{z}_1 & B_2 ^\dagger - \bar{z}_2 & 0 }
\pmatrix{\psi_1 \cr \psi_2 \cr \xi}= 0.
\end{equation}

For $k=1$ we can first choose $B_1=B_2=0$ by translation and we
get $I=\sqrt{\zeta} $ from the ADHM equation \eq{ADHM1}. Then
Eq.\eq{Dpsi} is solved as \be \l{zeromode1} \psi_1 =
\bar{z}_2\delta^{-1}I\xi, \quad \psi_2 = \bar{z}_1\delta^{-1}I\xi,
\ee where $\delta \equiv  x^2= z_1\bar{z}_1+ z_2\bar{z}_2$ and
$\xi = (1 + I^\dagger \delta^{-1} I)^{-\frac12}$. As emphasized
by Furuuchi \ct{KF} and Ho \ct{Ho}, note that the operator $\psi$
annihilates $|0,0\rangle$ state, so the projective module is
normalized as \be \l{p1} \psi^\dagger \psi=p\equiv
1-|0,0\rangle\langle0,0|. \ee If $\bar{z}_1,\,\bar{z}_2$ in
\eq{zeromode1} are ordered to the right of $\delta^{-1}I\xi$
according to the commutation rules \be \l{crule} \bar{z}_\a
f(\d)= f(\d+\zeta/2)\bar{z}_\a, \quad z_\a f(\d)= f(\d-\zeta/2)
z_\a, \;\;\;(\a=1,2) \ee for a function $f(z,\bar{z})$, the
$|0,0\rangle$ state is projected out from the Fock space and
$\d^{-1} \rightarrow (\d+\zeta/2)^{-1}$. So the ADHM solution
\eq{zeromode1} is well defined for all states in $\ch$. It is
crucial to prove the completeness relation \eq{CR} to observe that
ADHM always arrange their solutions to be singularity-free.

Now we can calculate the connection $A=\psi^\dagger d \psi$ in terms of
one variable $\xi$
\be
\l{A1}
A = \xi^{-1}\alpha\xi + \xi^{-1}d\xi,
\ee
where $\alpha= \xi^2 \pa_z \xi^{-2}
= - {\zeta \over \left(x^2 +{\zeta \over 2}\right)
\left(x^2 +\zeta\right)}\bar{z}_\a dz_\a$
and $\partial_z= dz_\alpha {\partial \over \partial z _\alpha},\;
\bar{\partial}_{\bar z}= d\bar{z}_\alpha {\partial \over \partial
\bar{z}_\alpha} \;(d=\partial_z+\bar{\partial}_{\bar z})$.
Then the corresponding field strength $F_A$ is obtained from
\begin{eqnarray}
\label{F}
  F_A &=& dA + A^2, \n\\
    &=& \xi^{-1}(d\alpha + \alpha^2 )\xi.
\end{eqnarray}
Even if we deal with the $U(1)$ case we have to keep the second
term in \eq{F} because of the gauge covariance on noncommutative
space. The field strength $F_A$ can be obtained from
\eq{Fformular} or by direct calculation of \eq{F} with attention
to ordering:
\begin{eqnarray}
\label{U1F}
&&F_A ={\zeta\over{x^2\left(x^2 +{\zeta\over 2}\right)
\left(x^2 +\zeta \right)}} \biggl(f_3( dz_2 d\bar{z}_2
- dz_1d\bar{z}_1 ) + f_+ d\bar{z}_1 dz_2
+ f_- d\bar{z}_2 dz_1\biggr), \nonumber \\
&& f_3 =  z_1 \bar{z}_1 - z_2 \bar{z}_2,\ \ f_+ = 2 z_1\bar{z}_2, \ \
f_-= 2 z_2 \bar{z}_1.
\end{eqnarray}
Since the ``origin'', i.e. $|0,0\rangle$, is projected out,
$F_A$ has no singularity.
The topological action density is given by
\begin{equation}
\label{Actden}
\hat{S} = - {1 \over 8{\pi^2}}F_A F_A =- {\zeta^2 \over {\pi^2}}
{1\over{x^2 \left({x^2+{\zeta\over2}}\right)^2
\left({x^2 + \zeta}\right)}}p,
\end{equation}
where we used the fact that $ dz_1\wedge d\bar{z}_1 \wedge
dz_2\wedge d\bar{z}_2=-4\,({\rm volume\; form})$. Note that
$\hat{S}$ for $k=1$ depends only on $x^2$ due to the rotational
symmetry.

The total action over noncommutative $\4r$ is defined by using the
prescription \eq{trace} as
\begin{equation}
\label{action2}
{\rm Tr}_\ch \hat{S} =\Bigl(\frac{\zeta\pi}{2}\Bigr)^2
\sum_{(n_1, n_2)\ne (0,0)}^{\infty} \hat{S}.
\end{equation}
Using the facts
\bea
\l{formular}
&&x^2 |n_1 , n_2\rangle =\frac\zeta2 (n_1 + n_2)|n_1 , n_2\rangle,\n\\
&&\sum_{(n_1, n_2)\ne (0,0)}^{\infty}\langle N|{\cal O}(x) |N
\rangle= \sum_{N=1}^{\infty}(N+1)\langle N|{\cal O}(x) |N
\rangle, \eea the instanton number for the solution \eq{A1} turns
out to be $-1$, that is, \be {\rm Tr}_\ch \hat{S}
 = - 4 \sum_{N=1}^{\infty} {1 \over N(N+1)(N+2)} =-1.
\ee

\subsection{Two $U(1)$ Instantons}

Next we will perform the same calculation for $U(1)$ instanton
solution with $k=2$. We start with the matrices satisfying the
ADHM constraints \eq{ADHM1} and \eq{ADHM2}:
\begin{equation}
B_1 = \pmatrix { 0 & \sqrt{\zeta} \cr 0 & 0 },
\quad B_2 = 0,
\quad  I = \pmatrix { 0 \cr \sqrt{2\zeta} },
\quad J =0,
\end{equation}
where we have fixed the moduli corresponding to the relative
position between two instantons for simplicity (for a general
solution with the moduli dependence, see \ct{KF}). With these
data we can get the normalized solution of Eq.\eq{Firstadhm}
\begin{eqnarray}
\l{k2}
&&\psi = \pmatrix {\psi_1 \cr \psi_2 \cr \xi}, \n\\
&& \psi_1 = \sqrt{2\zeta} \pmatrix { \sqrt{\zeta} \bar{z}_1\bar{z}_2  \cr
\bar{z}_2 (x^2 + \frac\zeta 2)  }Q^{-1},
\;\;\;\psi_2 = \sqrt{2\zeta} \pmatrix { \sqrt{\zeta} \bar{z}_1\bar{z}_1 \cr
\bar{z}_1 (x^2-\frac\zeta  2)}Q^{-1},\\
&&\xi = \left[{ -\zeta z_1\bar{z}_1 + x^2(x^2+\frac12 \zeta) } \over
  { -\zeta z_1\bar{z}_1 + (x^2+\frac\zeta 2)(x^2+ 2\zeta) }
\right]^\frac12,\n
\end{eqnarray}
where
$$Q = \left[\{ -\zeta z_1\bar{z}_1 + x^2(x^2+\frac12
    \zeta) \} \{ -\zeta z_1\bar{z}_1 + (x^2+\frac\zeta 2)(x^2+ 2\zeta) \}
\right]^\frac12.$$ Note that the states $\vert 0,0 \rangle$ and
$\vert 1,0 \rangle $ are annihilated by all components of $\psi$
\ct{KF} (where the projected states in general depend on the
moduli entering in $B_1$ and $B_2$). The operator $\psi$ in
\eq{k2} is thus normalized as \be \l{pk2} \psi^\dagger
\psi=p\equiv 1-|0,0\rangle\langle0,0|-|1,0\rangle\langle 1,0|,
\ee so $Q^{-1}$ is well-defined.

With this solution, the field strength $F_A$ can be calculated
with careful ordering from the formula \eq{Fformular}
\begin{eqnarray}
\label{U1K2F}
&&F_A ={2\zeta \over Q } \biggl(f_3(dz_2 d\bar{z}_2
- dz_1 d\bar{z}_1) + f_+ d\bar{z}_1 dz_2
+ f_- d\bar{z}_2 dz_1\biggr), \nonumber \\
&& f_3 =
 { G_1 \over Q P_1  } z_1 \bar{z}_1
- { G_2 \over Q P_2 } z_2 \bar{z}_2,
\quad  f_+= {2(x^2+\frac\zeta 2) H_1 \over {Q_1 P_1}}
z_1 \bar{z}_2,
\quad f_-= {2(x^2 + \frac\zeta 2) H_2 \over {Q_2 P_2}}
z_2 \bar{z}_1,\nonumber \\
&& G_1 := {\zeta (3x^2 + \frac\zeta 2)(z_1 \bar{z}_1 - \frac\zeta 2 )
+ x^2 (x^2 - \frac\zeta 2)^2 }, \nonumber \\
&& G_2 := { \zeta (3x^2 + \frac52 \zeta ) z_1 \bar{z}_1
+ x^2 (x^2 + \frac\zeta 2)^2 }, \nonumber \\
&& H_1 := { 3\zeta ( z_1 \bar{z}_1- \frac\zeta 2)
+  x^2 (x^2 - \frac\zeta 2)  }, \nonumber \\
&& H_2 := { 3\zeta z_1 \bar{z}_1
+  x^2 (x^2 - \frac\zeta 2)  }, \nonumber \\
&& Q_1 := \Bigl[\{-\zeta (z_1 \bar{z}_1-\frac\zeta 2)
+ x^2(x^2+\frac\zeta2)\}\{-\zeta (z_1 \bar{z}_1-\frac\zeta 2)
+ (x^2+\frac\zeta 2)(x^2+ 2\zeta)\}\Bigr]^\frac12, \nonumber \\
&& Q_2 := \Bigl[\{-\zeta (z_1 \bar{z}_1+\frac\zeta 2)
+ x^2(x^2+\frac\zeta2)\}\{-\zeta (z_1 \bar{z}_1+\frac\zeta 2)
+ (x^2+\frac\zeta 2)(x^2+ 2\zeta)\}\Bigr]^\frac12, \nonumber \\
&& P_1 := { -\zeta (z_1 \bar{z}_1 -\frac\zeta 2 )
+ x^2 ( x^2 + \frac32 \zeta) }, \nonumber \\
&& P_2 := { -\zeta z_1 \bar{z}_1
+ x^2 ( x^2 + \frac32 \zeta) }.
\end{eqnarray}
The $f_-$ component has a singularity coming from $Q_2$ at
$|0,1\rangle$ state which is not projected out by $p$. However,
$F_A$ is well-defined since the $|0,1\rangle$ state is annihilated
before it causes any trouble due to the factor $z_2 \bar{z}_1$ in
$f_-$. Notice that, in the case of $k=2$ instanton solution with
relative separation, we can not expect the spherical symmetry, so
the action depends on another coordinates in addition to $x^2$.

By straightforward calculation, the instanton charge density
${\hat S}$ can be explicitly calculated using the same
normalization that the $k=1$ case
\begin{eqnarray}
\label{K2action}
\hat{S} &=& - {1 \over 8{\pi^2}}F_A F_A \\
 &=& -{4\zeta^2 \over {\pi^2}} {\frac1{Q^2}}
\biggl[{1\over Q^2}\Bigl({G_1 \over P_1} z_1\bar{z}_1 -
{G_2 \over P_2} z_2 \bar{z}_2 \Bigr)^2
+ {{2(x^2+\frac\zeta 2)^2 H_1^2} \over {Q_1^2 P_1^2}}
z_1 \bar{z}_1 \bar{z}_2 z_2
+ {{2(x^2+\frac\zeta 2)^2 H_2^2}
\over {Q_2^2 P_2^2}} \bar{z}_1 z_1 z_2 \bar{z}_2 \biggr]p.\n
\end{eqnarray}
Now the instanton charge can be numerically calculated in the SHO
basis \eq{sho} (where the sum with respect to $n_1$ and $n_2$
should be separately done since the spherical symmetry is broken).
We performed this double infinite sum using Maple over $40,399$
points with $0\leq n_1 \leq 200, \;  0\leq n_2 \leq 200$
excluding the indicated points $(0,0),\;(1,0)$. The result is
\begin{equation}
\label{action2K2} {\rm Tr}_\ch \hat{S}
=\Bigl(\frac{\zeta\pi}{2}\Bigr)^2 \sum_{(n_1, n_2)\ne (0,0) \atop
(n_1, n_2)\ne(1,0)}^{\infty} \hat{S} \approx -{1.9998877}\approx
-2.
\end{equation}
Now this result is consistent with \ct{CKT}. \footnote{In the
previous version of this paper, we obtained ${\rm Tr}_\ch
\hat{S}=-0.932$ incorrectly due to an error of our numerical
calculation.} Following the argument in \ct{CKT}, we believe that
the instanton number for $U(1)$ solutions is always an integer,
independent of the moduli entering in $B_1,\,B_2$. This should be
the case since we have already introduced the integer number $k$
to specify the ADHM data.

\subsection{Single $U(2)$ Instanton}

Now we will seek for $U(2)$ solution \ct{KF} following the same steps
as the $U(1)$ case.
From the ADHM equations with $B_1=B_2=0$,
one can choose $I$ and $J$ as follows
\begin{eqnarray}
\l{IJ}
&&I = \pmatrix{{\sqrt{\rho^2+\zeta}} & 0 } := \pmatrix{a & 0},\nonumber \\
&&J = \pmatrix{ 0 \cr   \rho} := \pmatrix{0 \cr b},
\end{eqnarray}
where $\rho$ is a non-negative number and parameterizes the
classical size of the instanton. Then, from Eq.\eq{Firstadhm}, we
get the following solution
\begin{eqnarray}
&&\psi_1 = \bar{z}_2\delta^{-1}I\xi - z_1 \Delta^{-1}J^\dagger \xi,
\nonumber \\
&&\psi_2 = \bar{z}_1 \delta^{-1}I\xi + z_2 \Delta^{-1}J^\dagger \xi,\\
&&\xi = (1+I^{\dagger}\delta^{-1}I +
J\Delta^{-1}J^\dagger)^{-\frac12}, \nonumber
\end{eqnarray}
where $\Delta=\d + \zeta$. Using the explicit solution \eq{IJ},
$\xi$ is expressed as
\begin{eqnarray}
&&\xi = \pmatrix { \left({\d \over \Delta +\rho^2}\right)^\frac12   &  0 \cr
0 &\left({\Delta \over \Delta +\rho^2}\right)^\frac12  } :
=\pmatrix{\xi_- & 0 \cr 0 & \xi_+} \nonumber
\end{eqnarray}
and the zero-modes $\psi$ are
\begin{eqnarray}
\label{psi}
&&\psi_1 :=\pmatrix{f_1 & g_1} = \pmatrix{
\bar{z}_2({1 \over \d} a  \xi_-) & -z_1 ({1 \over \Delta} b\xi_+ )}, \n\\
&&\psi_2 :=\pmatrix{f_2 & g_2} = \pmatrix{\bar{z}_1 ({1 \over \d} a \xi_-)
& z_2 ({1 \over \Delta} b\xi_+ )}, \n\\
&& \psi := \pmatrix{\psi^{(1)} & \psi^{(2)}} = \pmatrix{f_1 & g_1 \cr
  f_2 & g_2 \cr \xi_- & 0 \cr 0 & \xi_+ }.
\end{eqnarray}
From the above expression we see that $\psi^{(1)}$ annihilates
the state $\left|0,0 \right\rangle$ for any values of $\rho$,
and normalized in the subspace where $\left|0,0 \right\rangle$ is
projected out, that is, $\psi^{(1)\dagger}\psi^{(1)}=p$.
The zero-mode $\psi^{(2)}$ annihilates no states in $\cal H$ and
manifestly nonsingular even if $\rho = 0$.
When $\rho = 0$, $g_1 = g_2 = 0$,
and, from (\ref{Fformular}), we see that
$\psi^{(2)}$ does not contribute
to the field strength.
Therefore the structure of the $U(2)$ instanton at $\rho = 0$
is completely determined by the minimal zero-mode $\psi^{(1)}$
in the $U(1)$ subgroup \ct{KF}.

The gauge field $A = \psi^\dagger d \psi$ can be now explicitly
calculated and the result is
\begin{eqnarray}
\l{A2}
&&A = \xi^{-1}\alpha\xi + \xi^{-1}d\xi, \n\\
&&\alpha = K \pmatrix { C_1 \bar{z}_\alpha dz_\alpha &
C_2 (z_1 dz_2-z_2 dz_1 )\cr
C_2 (\bar{z}_2 d\bar{z}_1-\bar{z}_1 d\bar{z}_2)
& C_3 z_\alpha d\bar{z}_\alpha  },
\end{eqnarray}
where
$$K ={1 \over (\d+{\zeta\over 2})(\Delta+\rho^2)},
\quad C_1=-(\rho^2+\zeta),
\quad C_2 = \rho \sqrt{\rho^2 + \zeta}, \quad C_3 = -\rho^2.$$
If we let $\zeta = 0$, we can get the ordinary $SU(2)$ instanton solution
\be
A_\mu = -2i\rho^2 \Sigma_{\mu\nu} {x_\nu \over x^2(x^2 +\rho^2) },
\ee
where $\Sigma_{\mu\nu}$ is the 't Hooft symbol
which is both antisymmetric and self-dual
with respect to their indices \ct{soin}.
On the other hand, if we let $\rho = 0$, we get
\begin{equation}
\alpha = - {\zeta \over \left(x^2 +{\zeta \over 2}\right)
\left(x^2 +\zeta\right)}
(\bar{z}_1 dz_1 + \bar{z}_2 dz_2),
\end{equation}
which is exactly equal to the $U(1)$ solution in \eq{A1} for the
reason explained above.

The field strength $F_A$ can be obtained from \eq{Fformular} or
by direct calculation with the solution \eq{A2} if one keeps in
mind careful ordering \footnote{Here we are using a shorthand
notation where $(\a\a)$ denotes the coordinates $z_\a z_\a$ and
$({\bar \a}\a)={\bar z}_\a z_\a$, etc.}:
\begin{eqnarray}
F_A=&&d\bar{z}_1 \wedge dz_1
\pmatrix{ \frac12 B_1( 2\bar{2}  - 1\bar{1}) &  B_2(12) \cr
 B_3 (\bar{1}\bar{2}) & \frac12 B_4(1\bar{1} - 2\bar{2})} \nonumber \\
&+&d\bar{z}_2 \wedge dz_2 \pmatrix{ \frac12 B_1( 1\bar{1} - 2\bar{2})
& - B_2(12) \cr
- B_3 (\bar{1}\bar{2}) & \frac12 B_4(2\bar{2} - 1\bar{1})} \nonumber \\
&+&d\bar{z}_1 \wedge dz_2 \pmatrix{-B_1(1\bar{2}) & -B_2(11) \cr
B_3(\bar{2}\bar{2}) & B_4(1\bar{2})} \nonumber \\
&+&dz_1 \wedge d\bar{z}_2 \pmatrix{B_1(\bar{1}2) & -B_2(22) \cr
B_3(\bar{1}\bar{1}) & -B_4(\bar{1}2)},
\end{eqnarray}
where
\begin{eqnarray}
&&B_1 := {2C_1 \over {\d(\d + \rho^2+\frac\zeta2 )
(\Delta+ \rho^2)}} \nonumber \\
&&B_2 := {2C_2 \over {\d(\d +\rho^2+ \frac\zeta2 )
(\Delta+ \rho^2)}}
\left(\Delta +\rho^2 \over \d+\rho^2 \right)^\frac12 \nonumber \\
&&B_3 := {2C_2 \over {\Delta(\Delta +   \rho^2)
(\Delta + \rho^2 +\frac\zeta2)}} \left(\Delta +\rho^2
\over \Delta +\rho^2+ \zeta\right)^\frac12 \nonumber  \\
&&B_4 := {2C_3 \over {\Delta(\Delta +   \rho^2)
(\Delta + \rho^2 +\frac\zeta2)}}. \nonumber
\end{eqnarray}
One can check that this $F_A$ is anti-Hermitian and
anti-self-dual $(F^A_{\bar{1}1}+F^A_{\bar{2}2}=0)$ using the rule
\eq{crule}. By straightforward calculation, one can determine the
instanton charge density which also depends only on $x^2$ due to
rotational symmetry
\begin{eqnarray}
\l{S2}
{\hat S}&&=-\frac{1}{8\pi^2}Tr (F_A \wedge F_A) \n\\
&&=-\frac{1}{2\pi^2}\Bigl\{(B_1^2+B_4^2)(1\bar{1}\bar{2}2+\bar{1}12\bar{2})
+B_2^2(11\bar{1}\bar{1}+22\bar{2}\bar{2})+
    B_3^2(\underbrace{\bar{1}\bar{1}11+\bar{2}\bar{2}22}) \Bigr\} \n\\
&&\quad -\frac{1}{4\pi^2}\Bigl\{(B_1^2+B_4^2)(1\bar{1}-2\bar{2})^2
+4 B_2^2 1\bar{1}2\bar{2}+4B_3^2\underbrace{\bar{1}1\bar{2}2}\Bigr\}\n\\
&&= -\frac{1}{4\pi^2}\Bigl\{(B_1^2 + B_4 ^2) x^2\left(x^2 + \zeta\right)
+2B_2^2 \Bigl( x^2 - {\zeta\over 2}\Bigr)x^2\Bigr\}p\nonumber \\
&&\quad -\frac{1}{2\pi^2}B_3^2\left( x^2 + \zeta\right)
\Bigl( x^2 + {3\zeta \over 2}\Bigr).
\end{eqnarray}
In the above expression, the parts except $\underbrace{( \cdots
)}$ project out the state $|0,0 \rangle$, so we explicitly
inserted the projection operator $p=1-|0,0 \rangle \langle0,0|$
in the parts. It can be confirmed again to recover the ordinary
$SU(2)$ instanton solution in the $\zeta = 0$ limit where
$B_1=\cdots=B_4$ and the $U(1)$ case for the limit $\rho =0$
where only $B_1$ term in \eq{S2} survives.

Finally we calculate the instanton charge of $U(2)$ solution. Note
that, since the part involved with $B_3^2$ in \eq{S2}, denoted as
${\hat S}_2$, does not project out any states in $\ch$, the trace
with respect to the part should be performed over the full
Hilbert space \eq{sho} including the state $|0,0 \rangle$, while
that involved with the projection operator $p$ in \eq{S2},
denoted as ${\hat S}_1$, has to exclude the ``origin'', $|0,0
\rangle$. Using the relations \eq{formular}, the topological
charge can be calculated (where we used Maple)
\begin{eqnarray}
\label{2instanton1}
{\rm Tr}_\ch \hat{S}
&=&\Bigl(\frac{\zeta\pi}{2}\Bigr)^2\Biggl( \sum_{N=1}^\infty
(N+1)\hat{S}_1(N) +\sum_{N=0}^\infty (N+1)\hat{S}_2(N)\Biggr)\\
&=&\sum_{N=1}^\infty \Biggl({4 \over
(N+1)(N+2)(N+2a^2)(N+2a^2+1)^2(N+2a^2+2)^2(N+2a^2+3)^2} \times
\nonumber\\
&&\Bigl((N^3+6N^2+11N+6)^2 + 2a^2(N+2)^3(3N^3+16N^2+25N+12) +
\nonumber\\
&& \;\;\; 2a^4(3N^6+45N^5+257N^4+714N^3+1028N^2+ 737N+212)+
\nonumber\\
&&\;\;\; 4a^6(9N^5+83N^4+301N^3+512N^2+ 400N+115) +
\nonumber\\
&&\;\;\; 8a^8 (9N^4+55N^3+122N^2+ 109N+30) + 16a^{10}(3N^3+11N^2+
12N+3)\Bigr)\Biggr) \nonumber \\
&=&-1,\nonumber
\end{eqnarray}
where $a =\rho/\sqrt{\zeta}$. Note that the dependence on the
instanton modulus $\rho$ remarkably dissappears in the final
answer. \footnote{Now this result is in agreement with the result
in \ct{CKT}. In the previous version of this paper, it was
incorrectly claimed due to a programming error that the instanton
number depends on the moduli.}

\section{Completeness Relation}

In this section we will show that our solutions we constructed in
section 3 exactly satisfy the completeness relation \eq{CR}. And
then it is shown that this completeness relation is a general
property satisfied in the ADHM construction.

The completeness relation \eq{CR} is actually a canonical
decomposition of a vector space ${\bf C}^{N+2k}$ (or a free
module $\ca^{\otimes N+2k}$ for noncommutative instantons) into
the null-space \eq{Firstadhm} and its orthogonal complement. This
decomposition is well-defined \ct{Ho} even in the noncommutative
space in spite of the nontrivial normalization \eq{p} since the
projective module (see footnote 1) corresponding to a vector
bundle is also well-defined in this case.

Let's start with the simplest case, the single $U(1)$ instanton in
section 3.1. In this case, the matrices $D f D^\dagger$ and $\psi
\psi^\dagger$ where $f^{-1}= z_1\bar{z}_1+ z_2\bar{z}_2 +\zeta
=\Delta$ have the forms
\begin{eqnarray}\label{Mat1}
&&D f D^\dagger =\pmatrix{
  z_1 f \bar{z}_1+ \bar{z}_2 f z_2
  & \bar{z}_2 f z_1-  z_1 f \bar{z}_2
  & -\sqrt{\zeta}\bar{z}_2 f \cr
  \bar{z}_1 f z_2-  z_2 f \bar{z}_1
   & \bar{z}_1 f z_1+ z_2 f \bar{z}_2
   &  -\sqrt{\zeta}\bar{z}_1 f \cr
    -\sqrt{\zeta}f z_2
   & -\sqrt{\zeta}f z_1  & \zeta f}\\
\label{Mat2} && \psi \psi^\dagger=\pmatrix{
  \bar{z}_2 {\zeta \over \delta\Delta} z_2
  & \bar{z}_2 {\zeta \over \delta\Delta} z_1
  & \bar{z}_2 {\sqrt{\zeta}\over \Delta} \cr
  \bar{z}_1 {\zeta \over \delta\Delta} z_2
  & \bar{z}_1 {\zeta \over \delta\Delta} z_1
  & \bar{z}_1 {\sqrt{\zeta}\over \Delta} \cr
  {\sqrt{\zeta}\over \Delta} z_2
  & {\sqrt{\zeta}\over \Delta} z_1
  & {\delta \over \Delta}}
\end{eqnarray}
To check the completeness relation \eq{CR} is now a simple
straightforward algebra using the formula \eq{crule}. Similarly
one can easily check the completeness relation for the $U(2)$
single instanton in section 3.3. For this case,
\begin{eqnarray}\label{Mat3}
&&D f D^\dagger =\pmatrix{
  z_1 f \bar{z}_1+ \bar{z}_2 f z_2
  & \bar{z}_2 f z_1-  z_1 f \bar{z}_2
  & -\sqrt{\rho^2+\zeta}\bar{z}_2 f
  & \rho z_1 f  \cr
  \bar{z}_1 f z_2-  z_2 f \bar{z}_1
   & \bar{z}_1 f z_1+ z_2 f \bar{z}_2
   &  -\sqrt{\rho^2+\zeta}\bar{z}_1 f
   & -\rho z_2 f \cr
    -\sqrt{\rho^2+\zeta}f z_2
   & -\sqrt{\rho^2+\zeta}f z_1  &(\rho^2+ \zeta)f & 0 \cr
   \rho f \bar{z}_1 & -\rho f \bar{z}_2
   & 0 & \rho^2 f}  \\
\label{Mat4} && \psi \psi^\dagger=\pmatrix{
  f_1 f_1^\dagger + g_1 g_1^\dagger
  & f_1 f_2^\dagger + g_1 g_2^\dagger
  & f_1 \xi_-^\dagger
  & g_1 \xi_+^\dagger \cr
  f_2 f_1^\dagger + g_2 g_1^\dagger
  & f_2 f_2^\dagger + g_2 g_2^\dagger
  & f_2 \xi_-^\dagger
  & g_1 \xi_+^\dagger \cr
   \xi_- f_1^\dagger
  & \xi_- f_2^\dagger
  & \xi_- \xi_-^\dagger & 0 \cr
  \xi_+ g_1^\dagger
  & \xi_+ g_2^\dagger  & 0
  & \xi_+ \xi_+^\dagger }
\end{eqnarray}
where $f^{-1}=z_1\bar{z}_1+ z_2\bar{z}_2 +\rho^2 +\zeta =\Delta
+\rho^2$ and the notations in \eq{Mat4} are coming from \eq{psi}.

Before checking the completeness for the case of the $U(1)$ two
instantons, let's argue that the completeness relation \eq{CR} in
the ADHM construction is a general property satisfied even for
noncommutative spaces. For this, it is important to observe the
following property
\begin{equation}\label{projection}
  \psi p =\psi,\;\;\; p \psi^\dagger =\psi^\dagger,
\end{equation}
where $p$ is the projection operator in \eq{p} which is
$p=1-|0,0\rangle\langle 0,0|$ for $k=1\;\; U(1)$ instanton and
$p=\pmatrix{1-|0,0\rangle\langle 0,0|& 0\cr 0 & 1}$ for
$k=1\;\;U(2)$ instanton. Of course the above property is quite
general in the ADHM construction for the reason stated below
\eq{crule}. When the null-space condition \eq{Firstadhm} is given
in a vector space ${\bf C}^{N+2k}$ or a free module $\ca^{\otimes
N+2k}$, one can ask whether the completeness relation \eq{CR} is
satisfied. If one notices the projection operators in left-hand
side of \eq{CR} are both well-defined, i.e. non-singular, this
relation should be satisfied since operating $D$ and $\psi$ from
the right-hand side or $D^\dagger$ and $\psi^\dagger$ from the
left-hand side the relation is always satisfied due to
\eq{Firstadhm} and \eq{projection} respectively. One can check
using the explicit expressions \eq{Mat1}-\eq{Mat4} that the
relation \eq{CR} is satisfied, as it should be, even for the
dangerous state $|0,0\rangle$. This general argument actually can
also be extracted from the construction of a projective instanton
module given by Ho \ct{Ho}. We checked this claim for the $U(1)$
two instantons in section 3.2 as well although a little but
straightforward algebra has been involved.

\section{Discussion}

We studied $U(1)$ and $U(2)$ instanton solutions on noncommutative
$\bf{R}^4$ based on the noncommutative version of ADHM equation
proposed by Nekrasov and Schwarz. It has been shown that the
anti-self-dual gauge fields on self-dual noncommutative $\4r$
correctly give integer instanton numbers for all cases we
consider.

We further showed that the completeness relation in the ADHM
construction is a general property satisfied even for
noncommutative spaces. To illustrate this claim more concretely,
let's consider the ADHM construction on ${\bf R}^2_{NC} \times
{\bf R}^2_{C}$ where ${\bf R}^2_{NC}$ is the noncommutative space
but ${\bf R}^2_{C}$ is the commutative space. This space is
represented by the algebra \be \label{nc2c2}
 [\bar{z}_1, z_1]=\zeta, \;\;\;
 [\bar{z}_2, z_2] =0. \ee
With this convention, one can easily check using the explicit
expressions \eq{Mat1}-\eq{Mat4} that the completeness relation is
exactly satisfied for this space. (Note that now
$z_2,\;\bar{z}_2$ are commutative coordinates, so one should apply
the formula \eq{crule} only for $\alpha=1$ with the change $\zeta
\rightarrow 2\zeta$.) Our result is different from \ct{CKT} by
Chu, et al. claiming that the completeness relation can be broken
down in this space. Indeed they argued that there is no
nonsingular $U(N)$ instanton on ${\bf R}^2_{NC} \times {\bf
R}^2_{C}$ due to the breakdown of the completeness relation. Our
present result may cure their ``unexpected" result correctly. We
hope to address this problem soon \ct{ourfc}.

We would like to mention some difference on finding an ADHM
solution between ours and \ct{CKT} although both methods
consequently give equivalent results. We have taken the
normalization \eq{p} throughout this paper and found the solutions
\eq{zeromode1}, \eq{k2} and \eq{psi} satisfying this
normalization. Alternatively one can take an another nomalization
$\psi^\dagger \psi =1$ like as \ct{CKT}. In this case one should
find the ADHM solution satisfying this normalization. Explicit
solutions of this kind were found in \ct{CKT}. However, the choice
of normalization is not important if and only if the consistent
solution could be found according to each normalization. In
addition, as shown in section 4, our normalization \eq{p} is
completely consistent with the completeness relation due to the
property \eq{projection} (and more economical on calculational
side).

\vspace{.5cm}
\noindent
{\large\bf Acknowledgments}
\\

We would like to thank Pei-Ming Ho for useful discussions and
comments and Koji Hashimoto, Eunsang Kim, and Satoshi Watamura
for helpful discussions on related matter. We thank C.-S. Chu, V.
V. Khoze and G. Travaglini for their critical comments in
\ct{CKT}. We were supported by the Ministry of Education, BK21
Project No. D-0055 and by grant No. 1999-2-112-001-5 from the
Interdisciplinary Research Program of the KOSEF. H.S.Y. was also
supported in part by Basic Science Research Institute in Sogang
University.

\newpage
\newcommand{\J}[4]{{ #1} {\bf #2} (#3) #4}
\newcommand{\andJ}[3]{{\bf #1} (#2) #3}
\newcommand{\AP}{Ann.\ Phys.\ (N.Y.)}
\newcommand{\MPL}{Mod.\ Phys.\ Lett.}
\newcommand{\NP}{Nucl.\ Phys.}
\newcommand{\PL}{Phys.\ Lett.}
\newcommand{\PR}{Phys.\ Rev.}
\newcommand{\PRL}{Phys.\ Rev.\ Lett.}
\newcommand{\CMP}{Comm.\ Math.\ Phys.}
\newcommand{\JMP}{J.\ Math.\ Phys.}
\newcommand{\JHEP}{J.\ High \ Energy \ Phys.}
\newcommand{\PTP}{Prog.\ Theor.\ Phys.}
\newcommand{\ib}{{\it ibid.}}
\newcommand{\hep}[1]{{\tt hep-th/{#1}}}


\end{document}